\documentclass[12pt]{article}

\usepackage{amsmath}
\usepackage{amssymb}
\usepackage{graphics}

\newcommand{\ip}[2]{\langle #1, #2 \rangle}

\title{\bf \Large Solving Non-uniqueness in Agglomerative Hierarchical Clustering Using Multidendrograms}
\author{Alberto Fern{\'a}ndez and Sergio G{\'o}mez\\
\\
\normalsize{Departament d'Enginyeria Inform\`{a}tica i Matem\`{a}tiques,}\\
\normalsize{Universitat Rovira i Virgili,}\\
\normalsize{Avinguda Pa\"{\i}sos Catalans 26,}\\
\normalsize{43007 Tarragona, Spain}\\
\\
\normalsize{E-mail: {\tt sergio.gomez@urv.cat}}\\
\normalsize{Software: {\tt http://deim.urv.cat/$\sim$sgomez/multidendrograms.php}}\\
\\
\normalsize{{\em Journal of Classification}, {\bf 25} (2008) 43-65}
}
\date{}

\begin{document}

\maketitle



\begin{quote}
\textbf{Abstract:} In agglomerative hierarchical clustering, pair-group methods suffer from a problem of non-uniqueness when two or more distances between different clusters coincide during the amalgamation process. The traditional approach for solving this drawback has been to take any arbitrary criterion in order to break ties between distances, which results in different hierarchical classifications depending on the criterion followed. In this article we propose a variable-group algorithm that consists in grouping more than two clusters at the same time when ties occur. We give a tree representation for the results of the algorithm, which we call a \textit{multidendrogram}, as well as a generalization of the Lance and Williams' formula which enables the implementation of the algorithm in a recursive way.
\end{quote}

\begin{quote}
\end{quote}

\section{Introduction}
Clustering methods group individuals into groups of individuals or \textit{clusters}, so that individuals in a cluster are close to one another. In agglomerative hierarchical clustering (Cormack 1971; Sneath and Sokal 1973, sec.~5.5; Gordon 1999, chap.~4), one begins with a proximity matrix between individuals, each one forming a singleton cluster. Then, clusters are themselves grouped into groups of clusters or \textit{superclusters}, the process being repeated until a complete hierarchy is formed. Among the different types of agglomerative methods we find single linkage, complete linkage, unweighted average, weighted average, etc., which differ in the definition of the proximity measure between clusters.

Except for the single linkage case, all the other clustering techniques suffer from a non-uniqueness problem, sometimes called the \textit{ties in proximity} problem, which is caused by ties either occurring in the initial proximity data or arising during the amalgamation process. From the family of agglomerative hierarchical methods, complete linkage is more susceptible than other methods to encounter ties during the clustering process, since it does not originate new proximity values different from the initial ones. With regard to the presence of ties in the original data, they are more frequent when one works with binary variables, or even with integer variables comprising just some few distinct values. But they can also appear using continuous variables, specially if the precision of experimental data is low. Sometimes, on the contrary, the absence of ties might be due to the representation of data with more decimal digits than it should be done. The non-uniqueness problem also depends on the measure used to obtain the proximity values from the initial variables. Moreover, in general, the larger the data set, the more ties arise (MacCuish, Nicolaou and MacCuish 2001).

The ties in proximity problem is well-known from several studies in different fields, for example in biology (Hart 1983; Backeljau, De Bruyn, De Wolf, Jordaens, Van Dongen and Winnepenninckx 1996; Arnau, Mars and Mar\'in 2005), in psychology (Van der Kloot, Spaans and Heiser 2005), or in chemistry (MacCuish et~al.\ 2001). Nevertheless, this problem is frequently ignored in software packages (Morgan and Ray 1995; Backeljau et~al.\ 1996; Van der Kloot et~al.\ 2005), and those packages which do not ignore it fail to adopt a common standard with respect to ties. Many of them simply break the ties in any arbitrary way, thus producing a single hierarchy. In some cases the analysis is repeated a given number of times with randomized input data order, and then additional criteria can be used for selecting one of the possible solutions (Arnau et~al.\ 2005). In other cases, some requirements are given on the number of individuals and the number of characteristics needed to generate proximity data without ties (Hart 1983; MacCuish et~al.\ 2001). None of these proposals can ensure the complete absence of ties, neither can all their requirements be satisfied always.

Another possibility for dealing with multiple solutions is to use further criteria, like a distortion measure (Cormack 1971, table~3), and select the best solution among all the possible ones. However, the result of this approach will depend on the distortion measure used, which means that an additional choice must be made. But this proposal does not ensure the uniqueness of the solution, since several candidate solutions might share the same minimum distortion value. Besides, in ill conditioned problems (those susceptible to the occurrence of too many ties), it is not feasible to perform an exhaustive search for all possible hierarchical classifications, due to its high computational cost. With regard to this, Van der Kloot et~al.\ (2005) analyze two data sets using many random permutations of the input data order, and with additional criteria they evaluate the quality of each solution. They show that the best solutions frequently emerge after many permutations, and they also notice that the goodness of these solutions necessarily depends on the number of permutations used.

An alternative proposal is to seek a hierarchical classification which describes common structure among all the possible solutions, as recommended by Hart (1983). One approach is to prune as little as possible from the classifications being compared to arrive at a common structure such as the maximal common pruned tree (Morgan and Ray 1995). Care must be taken not to prune too much, so this approach can be followed only when the number of alternative solutions is small and they are all known. Furthermore, the maximal common pruned tree need not be uniquely defined and it does not give a complete classification for all the individuals under study.

What we propose in this article is an agglomerative hierarchical clustering algorithm that solves the ties in proximity problem by merging into the same supercluster all the clusters that fall into a tie. In order to do so we must be able to calculate the distance separating any two superclusters, hence we have generalized the definition of distance between clusters to the superclusters case, for the most commonly used agglomerative hierarchical clustering techniques. Additionally, we give the corresponding generalization of Lance and Williams' formula, which enables us to compute these distances in a recursive way. Moreover, we introduce a new tree representation for the results obtained with the agglomerative algorithm: the multidendrogram.

In Section~2 we introduce our proposal of clustering algorithm and the multidendrogram representation for the results. Section~3 gives the corresponding generalization of some hierarchical clustering strategies. In Section~4, Lance and Williams' formula is also generalized consistently with the new proposal. Section~5 shows some results corresponding to data from a real example, and we finish with some conclusions in Section~6.

\section{Agglomerative Hierarchical Algorithm}
\subsection{Pair-Group Approach}
Agglomerative hierarchical procedures build a hierarchical classification in a bottom-up way, from a proximity matrix containing dissimilarity data between individuals of a set $\Omega = \{x_{1},x_{2},\ldots,x_{n}\}$ (the same analysis could be done using similarity data). The algorithm has the following steps:
\begin{enumerate}
  \item[0)] Initialize $n$ singleton clusters with one individual in each of them: $\{x_{1}\}$, $\{x_{2}\}$, \ldots, $\{x_{n}\}$. Initialize also the distances between clusters, $D(\{x_{i}\},\{x_{j}\})$, with the values of the distances between individuals, $d(x_{i},x_{j})$:
    \begin{displaymath}
      D(\{x_{i}\},\{x_{j}\}) = d(x_{i},x_{j}) \qquad \forall i,j=1,2,\ldots,n.
    \end{displaymath}
  \item[1)] Find the shortest distance separating two different clusters.
  \item[2)] Select two clusters $X_{i}$ and $X_{i'}$ separated by such shortest distance and merge them into a new supercluster $X_{i} \cup X_{i'}$.
  \item[3)] Compute the distances $D(X_{i} \cup X_{i'},X_{j})$ between the new supercluster $X_{i} \cup X_{i'}$ and each of the other clusters $X_{j}$.
  \item[4)] If all individuals are not in a single cluster yet, then go back to step~1.
\end{enumerate}

Following Sneath and Sokal (1973, p.~216), this type of approach is known as a \textit{pair-group} method, in opposition to \textit{variable-group} methods which will be discussed in the next subsection. Depending on the criterion used for the calculation of distances in step~3, we can implement different agglomerative hierarchical methods. In this article we study some of the most commonly used ones, which are: single linkage, complete linkage, unweighted average, weighted average, unweighted centroid, weighted centroid and joint between-within. The problem of non-uniqueness may arise at step~2 of the algorithm, when two or more pairs of clusters are separated by the shortest distance value (i.e., the shortest distance is tied). Every choice for breaking ties may have important consequences, because it changes the collection of clusters and the distances between them, possibly resulting in different hierarchical classifications. It must be noted here that not all tied distances will produce ambiguity: they have to be the shortest ones and they also have to involve a common cluster. On the other hand, ambiguity is not limited to cases with ties in the original proximity values, but ties may arise during the clustering process too.

The use of any hierarchical clustering technique on a finite set $\Omega$ with $n$ individuals results in an $n$\textit{-tree} on $\Omega$, which is defined as a subset $T$ of parts of $\Omega$ satisfying the following conditions:
\begin{enumerate}
  \item[(i)]   $\Omega \in T$,
  \item[(ii)]  $\emptyset \notin T$,
  \item[(iii)] $\forall x \in \Omega \quad \{x\} \in T$,
  \item[(iv)]  $\forall X,Y \in T \quad (X \cap Y = \emptyset \quad \vee \quad X \subseteq Y \quad \vee \quad Y \subseteq X)$.
\end{enumerate}
An $n$-tree gives only the hierarchical structure of a classification, but the use of a hierarchical clustering technique also associates a height $h$ with each of the clusters obtained. All this information is gathered in the definition of a \textit{valued tree} on $\Omega$, which is a pair $(T,h)$ where $T$ is an $n$-tree on $\Omega$ and $h:T \longrightarrow \mathbb{R}$ is a function such that $\forall X,Y \in T$:
\begin{enumerate}
  \item[(i)]   $h(X) \geq 0$,
  \item[(ii)]  $h(X)=0 \quad \Longleftrightarrow \quad |X|=1$,
  \item[(iii)] $X \subsetneq Y \quad \Longrightarrow \quad h(X)<h(Y)$,
\end{enumerate}
where $|X|$ denotes the cardinality of $X$.

For example, suppose that we have a graph with four individuals like that of Figure~1, where the initial distance between any two individuals is the value of the shortest path connecting them. This means, for example, that the initial distance between $x_{2}$ and $x_{4}$ is equal to~5. Using the unweighted average criterion, we can obtain three different valued trees. The graphical representation of valued trees are the so called \textit{dendrograms}, and Figure~2 shows the three corresponding dendrograms obtained for our toy graph. The first two dendrograms are quite similar, but the third one shows a considerably different hierarchical structure. Hence, if the third dendrogram is the only one obtained by a software package, one could extract from it the wrong conclusion that $x_{3}$ is closer to $x_{4}$ than it is to $x_{2}$.

\subsection{Variable-Group Proposal: Multidendrograms}
Any decision taken to break ties in the toy graph of Figure~1 would be arbitrary. In fact, the use of an unfortunate rule might lead us to the worst dendrogram of the three. A logical solution to the pair-group criterion problem might be to assign the same importance to all tied distances and, therefore, to use a variable-group criterion. In our example of Figure~1 this means the amalgamation of individuals $x_{1}$, $x_{2}$ and $x_{3}$ in a single cluster at the same time. The immediate consequence is that we have to calculate the distance between the new cluster $\{x_{1}\} \cup \{x_{2}\} \cup \{x_{3}\}$ and the cluster $\{x_{4}\}$. In the unweighted average case this distance is equal to~5, that is, the arithmetic mean among the values~7, 5 and 3, corresponding respectively to the distances $D(\{x_{1}\},\{x_{4}\})$, $D(\{x_{2}\},\{x_{4}\})$ and $D(\{x_{3}\},\{x_{4}\})$. We must also decide what height should be assigned to the new cluster formed by $x_{1}$, $x_{2}$ and $x_{3}$, which could be any value between the minimum and the maximum distances that separate any two of them. In this case the minimum distance is~2 and corresponds to both of the tied distances $D(\{x_{1}\},\{x_{2}\})$ and $D(\{x_{2}\},\{x_{3}\})$, while the maximum distance is the one separating $x_{1}$ from $x_{3}$ and it is equal to~4.

Following the variable-group criterion on a finite set $\Omega$ with $n$ individuals, we no longer get several valued trees, but we obtain a unique tree which we call a \textit{multivalued tree} on $\Omega$, and we define it as a triplet $(T,h_{l},h_{u})$ where $T$ is an $n$-tree on $\Omega$ and $h_{l},h_{u}:T \longrightarrow \mathbb{R}$ are two functions such that $\forall X,Y \in T$:
\begin{enumerate}
  \item[(i)]   $0 \leq h_{l}(X) \leq h_{u}(X)$,
  \item[(ii)]  $h_{l}(X)=0 \quad \Longleftrightarrow \quad h_{u}(X)=0 \quad \Longleftrightarrow \quad |X|=1$,
  \item[(iii)] $X \subsetneq Y \quad \Longrightarrow \quad h_{l}(X)<h_{l}(Y)$.
\end{enumerate}
A multivalued tree associates with every cluster $X$ in the hierarchical classification two height values, $h_{l}(X)$ and $h_{u}(X)$,  corresponding respectively to the lower and upper bounds at which member individuals can be merged into cluster $X$. When $h_{l}(X)$ and $h_{u}(X)$ coincide for every cluster $X$, the multivalued tree is just a valued tree. But, when there is any cluster $X$ for which $h_{l}(X) < h_{u}(X)$, it is like having multiple valued trees because every selection of a height $h(X)$ inside the interval $[h_{l}(X),h_{u}(X)]$ corresponds to a different valued tree. The length of the interval indicates the degree of heterogeneity inside cluster $X$. We also introduce here the concept of \textit{multidendrogram} to refer to the graphical representation of a multivalued tree. In Figure~3 we show the corresponding multidendrogram for the toy example. The shadowed region between heights 2 and 4 refers to the interval between the respective values of $h_{l}$ and $h_{u}$ for cluster $\{x_{1}\} \cup \{x_{2}\} \cup \{x_{3}\}$, which in turn also correspond to the minimum and maximum distances separating any two of the constituent clusters $\{x_{1}\}$, $\{x_{2}\}$ and $\{x_{3}\}$.

Let us consider the situation shown in Figure~4, where nine different clusters are to be grouped into superclusters. The clusters to be amalgamated should be those separated by the shortest distance. The picture shows the edges connecting clusters separated by such shortest distance, so we observe that there are six pairs of clusters separated by shortest edges. A pair-group clustering algorithm typically would select any of these pairs, for instance $(X_{8},X_{9})$, and then it would compute the distance between the new supercluster $X_{8} \cup X_{9}$ and the rest of the clusters $X_{i}$, for all $i \in \{1,2,\ldots,7\}$. What we propose here is to follow a variable-group criterion and create as many superclusters as groups of clusters connected by shortest edges. In Figure~4, for instance, the nine initial clusters would be grouped into the four following superclusters: $X_{1}$, $X_{2} \cup X_{3}$, $X_{4} \cup X_{5} \cup X_{6}$ and $X_{7} \cup X_{8} \cup X_{9}$. Then, all the pairwise distances between the four superclusters should be computed. In general, we must be able to compute distances $D(X_{I},X_{J})$ between any two superclusters $X_{I}=\bigcup_{i \in I}X_{i}$ and $X_{J}=\bigcup_{j \in J}X_{j}$, each one of them made up of several clusters indexed by $I=\{i_{1},i_{2},\ldots,i_{p}\}$ and $J=\{j_{1},j_{2},\ldots,j_{q}\}$, respectively.

The algorithm that we propose in order to ensure uniqueness in agglomerative hierarchical clustering has the following steps:
\begin{enumerate}
  \item[0)] Initialize $n$ singleton clusters with one individual in each of them: $\{x_{1}\}$, $\{x_{2}\}$, \ldots, $\{x_{n}\}$. Initialize also the distances between clusters, $D(\{x_{i}\},\{x_{j}\})$, with the values of the distances between individuals, $d(x_{i},x_{j})$:
    \begin{displaymath}
      D(\{x_{i}\},\{x_{j}\}) = d(x_{i},x_{j}) \qquad \forall i,j=1,2,\ldots,n.
    \end{displaymath}
  \item[1)] Find the shortest distance separating two different clusters, and record it as $D_{lower}$.
  \item[2)] Select all the groups of clusters separated by shortest distance $D_{lower}$ and merge them into several new superclusters $X_{I}$. The result of this step can be some superclusters made up of just one single cluster ($|I|=1$), as well as some superclusters made up of various clusters ($|I|>1$). Notice that the latter superclusters all must satisfy the condition $D_{min}(X_{I}) = D_{lower}$, where
    \begin{displaymath}
      D_{min}(X_{I}) = \min_{i \in I} \, \min_{\substack{i' \in I \\ i' \not = i}} \, D(X_{i},X_{i'}).
    \end{displaymath}
  \item[3)] Update the distances between clusters following the next substeps:
    \begin{enumerate}
      \item[3.1)] Compute the distances $D(X_{I},X_{J})$ between all superclusters, and record the minimum of them as $D_{next}$ (this will be the shortest distance $D_{lower}$ in the next iteration of the algorithm).
      \item[3.2)] For each supercluster $X_{I}$ made up of various clusters ($|I|>1$), assign a common amalgamation interval $[D_{lower},D_{upper}]$ for all its constituent clusters $X_{i}$, $i \in I$, where $D_{upper} = D_{max}(X_{I})$ and
        \begin{displaymath}
          D_{max}(X_{I}) = \max_{i \in I} \, \max_{\substack{i' \in I \\ i' \not = i}} \, D(X_{i},X_{i'}).
        \end{displaymath}
    \end{enumerate}
  \item[4)] If all individuals are not in a single cluster yet, then go back to step~1.
\end{enumerate}

Using the pair-group algorithm, only the centroid methods (weighted and unweighted) may produce \textit{reversals}. Let us remember that a reversal arises in a valued tree when it contains at least two clusters $X$ and $Y$ for which $X \subset Y$ but $h(X)>h(Y)$ (Morgan and Ray 1995). In the case of the variable-group algorithm, reversals may appear in substep~3.2. Although reversals make dendrograms difficult to interpret if they occur during the last stages of the agglomeration process, it can be argued that they are not very disturbing if they occur during the first stages. Thus, as happens with the centroid methods in the pair-group case, it could be reasonable to use the variable-group algorithm as long as no reversals at all or only unimportant ones were produced.

Sometimes, in substep~3.2 of the variable-group clustering algorithm, it will not be enough to adopt a fusion interval, but it will be necessary to obtain an exact fusion value (e.g., in order to calculate a distortion measure). In these cases, given the lower and upper bounds at which the tied clusters can merge into a supercluster, one possibility is to select the fusion value naturally suggested by the method being applied. For instance, in the case of the toy example and the corresponding multidendrogram shown in Figures~1 and~3, the fusion value would be $2.7$ (the unweighted average distance). If the clustering method used was a different one such as single linkage or complete linkage, then the fusion value would be $2$ or $4$, respectively. Another possibility is to use systematically the shortest distance as the fusion value for the tied clusters. Both criteria allow the recovering of the pair-group result for the single linkage method. The latter criterion, in addition, avoids the appearance of reversals. However, it must be emphasized that the adoption of exact fusion values, without considering the fusion intervals at their whole lengths, means that some valuable information regarding the heterogeneity of the clusters is being lost.

\section{Generalization of Agglomerative Hierarchical Methods}
In the variable-group clustering algorithm previously proposed we have seen the necessity of agglomerating simultaneously two families of clusters, respectively indexed by $I=\{i_{1},i_{2},\ldots,i_{p}\}$ and $J=\{j_{1},j_{2},\ldots,j_{q}\}$, into two superclusters $X_{I}=\bigcup_{i \in I}X_{i}$ and $X_{J}=\bigcup_{j \in J}X_{j}$. In the following subsections we derive, for each of the most commonly used agglomerative hierarchical clustering strategies, the distance between the two superclusters, $D(X_{I},X_{J})$, in terms of the distances between the respective component clusters, $D(X_{i},X_{j})$.

\subsection{Single Linkage}
In \textit{single linkage} clustering, also called \textit{nearest neighbor} or \textit{minimum} method, the distance between two clusters $X_{i}$ and $X_{j}$ is defined as the distance between the closest pair of individuals, one in each cluster:
\begin{equation}
  \label{single1}
  D(X_{i},X_{j}) = \min_{x \in X_{i}} \, \min_{y \in X_{j}} \, d(x,y).
\end{equation}
This means that the distance between two superclusters $X_{I}$ and $X_{J}$ can be defined as
\begin{equation}
  \label{single2}
  D(X_{I},X_{J}) = \min_{x \in X_{I}} \, \min_{y \in X_{J}} \, d(x,y) = \min_{i \in I} \, \min_{x \in X_{i}} \, \min_{j \in J} \, \min_{y \in X_{j}} \, d(x,y).
\end{equation}
Notice that this formulation generalizes the definition of distance between clusters in the sense that equation~(\ref{single1}) is recovered from equation~(\ref{single2}) when $|I|=|J|=1$, that is, when superclusters $I$ and $J$ are both composed of a single cluster. Grouping terms and using the definition in equation~(\ref{single1}), we get the equivalent definition:
\begin{equation}
  \label{single3}
  D(X_{I},X_{J}) = \min_{i \in I} \, \min_{j \in J} \, D(X_{i},X_{j}).
\end{equation}

\subsection{Complete Linkage}
In \textit{complete linkage} clustering, also known as \textit{furthest neighbor} or \textit{maximum} method, cluster distance is defined as the distance between the most remote pair of individuals, one in each cluster:
\begin{equation}
  \label{complete1}
  D(X_{i},X_{j}) = \max_{x \in X_{i}} \, \max_{y \in X_{j}} \, d(x,y).
\end{equation}
Starting from equation~(\ref{complete1}) and following the same reasoning as in the single linkage case, we extend the definition of distance to the superclusters case as
\begin{equation}
  \label{complete2}
  D(X_{I},X_{J}) = \max_{i \in I} \, \max_{j \in J} \, D(X_{i},X_{j}).
\end{equation}

\subsection{Unweighted Average}
\textit{Unweighted average} clustering, also known as \textit{group average} method or \textit{UPGMA} (Unweighted Pair-Group Method using Averages), iteratively forms clusters made up of pairs of previously formed clusters, based on the arithmetic mean distances between their member individuals. It uses an unweighted averaging procedure, that is, when clusters are joined to form a larger cluster, the distance between this new cluster and any other cluster is calculated weighting each individual in those clusters equally, regardless of the structural subdivision of the clusters:
\begin{equation}
  \label{average:unweighted1}
  D(X_{i},X_{j}) = \frac{1}{|X_{i}||X_{j}|} \sum_{x \in X_{i}} \sum_{y \in X_{j}} d(x,y).
\end{equation}
When the variable-group strategy is followed, the UPGMA name of the method should be modified to that of \textit{UVGMA} (Unweighted Variable-Group Method using Averages), and the distance definition between superclusters in this case should be
\begin{eqnarray*}
  \lefteqn{ D(X_{I},X_{J}) = \frac{1}{|X_{I}||X_{J}|} \sum_{x \in X_{I}} \sum_{y \in X_{J}} d(x,y) } \\
    & & = \frac{1}{|X_{I}||X_{J}|} \sum_{i \in I} \sum_{x \in X_{i}} \sum_{j \in J} \sum_{y \in X_{j}} d(x,y).
\end{eqnarray*}
Using equation~(\ref{average:unweighted1}), we get the desired definition in terms of the distances between component clusters:
\begin{equation}
  \label{average:unweighted2}
  D(X_{I},X_{J}) = \frac{1}{|X_{I}||X_{J}|} \sum_{i \in I} \sum_{j \in J} |X_{i}||X_{j}| D(X_{i},X_{j}).
\end{equation}
In this case, $|X_{I}|$ is the number of individuals in supercluster $X_{I}$, that is, $|X_{I}| = \sum_{i \in I} |X_{i}|$.

\subsection{Weighted Average}
In \textit{weighted average} strategy, also called \textit{WVGMA} (Weighted Variable-Group Method using Averages) in substitution of the corresponding pair-group name \textit{WPGMA}, we calculate the distance between two superclusters $X_{I}$ and $X_{J}$ by taking the arithmetic mean of the pairwise distances, not between individuals in the original matrix of distances, but between component clusters in the matrix used in the previous iteration of the procedure:
\begin{equation}
  \label{average:weighted}
  D(X_{I},X_{J}) = \frac{1}{|I||J|} \sum_{i \in I} \sum_{j \in J} D(X_{i},X_{j}).
\end{equation}

This method is related to the unweighted average one in that the former derives from the latter when we consider
\begin{equation}
  \label{relation}
  |X_{i}|=1 \quad \forall i \in I \qquad \mbox{and} \qquad |X_{j}|=1 \quad \forall j \in J.
\end{equation}
It weights the most recently admitted individuals in a cluster equally to its previous members. The weighting discussed here is with reference to individuals composing a cluster and not to the average distances in Lance and Williams' recursive formula (see next section), in which equal weights apply for weighted clustering and different weights apply for unweighted clustering (Sneath and Sokal 1973, p.~229).

\subsection{Unweighted Centroid}
The next three clustering techniques assume that individuals can be represented by points in Euclidean space. This method and the next one further assume that the measure of dissimilarity between any pair of individuals is the squared Euclidean distance between the corresponding pair of points. When the dissimilarity between two clusters $X_{i}$ and $X_{j}$ is defined to be the squared distance between their centroids, we are performing \textit{unweighted centroid} (or simply \textit{centroid}) clustering, also called \textit{UPGMC} (Unweighted Pair-Group Method using Centroids):
\begin{equation}
  \label{centroid:unweighted1}
  D(X_{i},X_{j}) = \|\overline{x}_{i} - \overline{x}_{j}\|^{2},
\end{equation}
where $\overline{x}_{i}$ and $\overline{x}_{j}$ are the centroids of the points in clusters $X_{i}$ and $X_{j}$ respectively, and $\| \cdot \|$ is the Euclidean norm. Therefore, under the variable-group point of view, the method could be named \textit{UVGMC} and the distance between two superclusters can be generalized to the definition:
\begin{equation}
  \label{centroid:unweighted2}
  D(X_{I},X_{J}) = \|\overline{x}_{I} - \overline{x}_{J}\|^{2}.
\end{equation}
In the Appendix it is proved that this definition can be expressed in terms of equation~(\ref{centroid:unweighted1}) as
\begin{eqnarray}
  \label{centroid:unweighted3}
  \lefteqn{ D(X_{I},X_{J}) = \frac{1}{|X_{I}||X_{J}|} \sum_{i \in I} \sum_{j \in J} |X_{i}||X_{j}| D(X_{i},X_{j}) } \nonumber \\
    & & - \frac{1}{|X_{I}|^{2}} \sum_{i \in I} \sum_{\substack{i' \in I \\ i'>i}} |X_{i}||X_{i'}| D(X_{i},X_{i'}) \nonumber \\
    & & - \frac{1}{|X_{J}|^{2}} \sum_{j \in J} \sum_{\substack{j' \in J \\ j'>j}} |X_{j}||X_{j'}| D(X_{j},X_{j'}).
\end{eqnarray}

\subsection{Weighted Centroid}
In \textit{weighted centroid} strategy, also called \textit{median} method or \textit{WVGMC} (Weighted Variable-Group Method using Centroids) in substitution of the pair-group name \textit{WPGMC}, we modify the definition of dissimilarity between two clusters given in the unweighted centroid case, assigning each cluster the same weight in calculating the ``centroid''. Now the center of a supercluster $X_{I}$ is the average of the centers of the constituent clusters:
\begin{displaymath}
  \overline{x}_{I} = \frac{1}{|I|} \sum_{i \in I} \overline{x}_{i}.
\end{displaymath}
This clustering method is related to the unweighted centroid one by relation~(\ref{relation}), which also related the weighted average strategy to the corresponding unweighted average. So, in this case we define the distance between two superclusters as
\begin{eqnarray}
  \label{centroid:weighted}
  \lefteqn{ D(X_{I},X_{J}) = \frac{1}{|I||J|} \sum_{i \in I} \sum_{j \in J} D(X_{i},X_{j}) } \nonumber \\
    & & - \frac{1}{|I|^{2}} \sum_{i \in I} \sum_{\substack{i' \in I \\ i'>i}} D(X_{i},X_{i'}) - \frac{1}{|J|^{2}} \sum_{j \in J} \sum_{\substack{j' \in J \\ j'>j}} D(X_{j},X_{j'}).
\end{eqnarray}

\subsection{Joint Between-Within}
Sz\'ekely and Rizzo (2005) propose an agglomerative hierarchical clustering method that minimizes a \textit{joint between-within} cluster distance, measuring both heterogeneity between clusters and homogeneity within clusters. This method extends Ward's minimum variance method (Ward 1963) by defining the distance between two clusters $X_{i}$ and $X_{j}$ in terms of any power $\alpha \in (0,2]$ of Euclidean distances between individuals:
\begin{eqnarray}
  \label{joint1}
  \lefteqn{ D(X_{i},X_{j}) = \frac{|X_{i}||X_{j}|}{|X_{i}|+|X_{j}|} \biggl( \frac{2}{|X_{i}||X_{j}|} \sum_{x \in X_{i}} \sum_{y \in X_{j}} \|x-y\|^{\alpha} } \nonumber \\
    & & - \frac{1}{|X_{i}|^{2}} \sum_{x \in X_{i}} \sum_{x' \in X_{i}} \|x-x'\|^{\alpha} - \frac{1}{|X_{j}|^{2}} \sum_{y \in X_{j}} \sum_{y' \in X_{j}} \|y-y'\|^{\alpha} \biggr).
\end{eqnarray}
When $\alpha = 2$, cluster distances are a weighted squared distance between cluster centers
\begin{displaymath}
  D(X_{i},X_{j}) = \frac{2|X_{i}||X_{j}|}{|X_{i}|+|X_{j}|} \|\overline{x}_{i} - \overline{x}_{j}\|^{2},
\end{displaymath}
equal to twice the cluster distance that is used in Ward's method.

In the Appendix we derive the following recursive formula for updating cluster distances as a generalization of equation~(\ref{joint1}):
\begin{eqnarray}
  \label{joint2}
  \lefteqn{ D(X_{I},X_{J}) = \frac{1}{|X_{I}|+|X_{J}|} \sum_{i \in I} \sum_{j \in J} (|X_{i}|+|X_{j}|) D(X_{i},X_{j}) } \nonumber \\
    & & - \frac{|X_{J}|}{|X_{I}|(|X_{I}|+|X_{J}|)} \sum_{i \in I} \sum_{\substack{i' \in I \\ i'>i}} (|X_{i}|+|X_{i'}|) D(X_{i},X_{i'}) \nonumber \\
    & & - \frac{|X_{I}|}{|X_{J}|(|X_{I}|+|X_{J}|)} \sum_{j \in J} \sum_{\substack{j' \in J \\ j'>j}} (|X_{j}|+|X_{j'}|) D(X_{j},X_{j'}).
\end{eqnarray}

\section{Generalization of Lance and Williams' Formula}
Lance and Williams (1966) put the most commonly used agglomerative hierarchical strategies into a single system, avoiding the necessity of a separate computer program for each of them. Assume three clusters $X_{i}$, $X_{i'}$ and $X_{j}$, containing $|X_{i}|$, $|X_{i'}|$ and $|X_{j}|$ individuals respectively and with distances between them already determined as $D(X_{i},X_{i'})$, $D(X_{i},X_{j})$ and $D(X_{i'},X_{j})$. Further assume that the smallest of all distances still to be considered is $D(X_{i},X_{i'})$, so that $X_{i}$ and $X_{i'}$ are joined to form a new supercluster $X_{i} \cup X_{i'}$, with $|X_{i}|+|X_{i'}|$ individuals. Lance and Williams express $D(X_{i} \cup X_{i'},X_{j})$ in terms of the distances already defined, all known at the moment of fusion, using the following recurrence relation:
\begin{eqnarray}
  \label{formula:Lance}
  \lefteqn{ D(X_{i} \cup X_{i'},X_{j}) = \alpha_{i} D(X_{i},X_{j}) + \alpha_{i'} D(X_{i'},X_{j}) } \nonumber \\
    & & + \, \beta D(X_{i},X_{i'}) + \gamma |D(X_{i},X_{j}) - D(X_{i'},X_{j})|.
\end{eqnarray}
With this technique superclusters can always be computed from previous clusters and it is not necessary to return to the original dissimilarity data during the clustering process. The values of the parameters $\alpha_{i}$, $\alpha_{i'}$, $\beta$ and $\gamma$ determine the nature of the sorting strategy. Table~1 gives the values of the parameters that define the most commonly used agglomerative hierarchical clustering methods.

We next give a generalization of formula~(\ref{formula:Lance}) compatible with the amalgamation of more than two clusters simultaneously. Suppose that one wants to agglomerate two superclusters $X_{I}$ and $X_{J}$, respectively indexed by $I=\{i_{1},i_{2},\ldots,i_{p}\}$ and $J=\{j_{1},j_{2},\ldots,j_{q}\}$. We define the distance between them as
\begin{eqnarray}
  \label{formula:new}
  \lefteqn{ D(X_{I},X_{J}) = \sum_{i \in I} \sum_{j \in J} \alpha_{ij} D(X_{i},X_{j}) } \nonumber \\
    & & + \sum_{i \in I} \sum_{\substack{i' \in I \\ i'>i}} \beta_{ii'} D(X_{i},X_{i'}) + \sum_{j \in J} \sum_{\substack{j' \in J \\ j'>j}} \beta_{jj'} D(X_{j},X_{j'}) \nonumber \\
    & & + \, \delta \sum_{i \in I} \sum_{j \in J} \gamma_{ij} [D_{max}(X_{I},X_{J}) - D(X_{i},X_{j})] \nonumber \\
    & & - (1 - \delta) \sum_{i \in I} \sum_{j \in J} \gamma_{ij} [D(X_{i},X_{j}) - D_{min}(X_{I},X_{J})],
\end{eqnarray}
where
\begin{displaymath}
  D_{max}(X_{I},X_{J}) = \max_{i \in I} \, \max_{j \in J} \, D(X_{i},X_{j})
\end{displaymath}
and
\begin{displaymath}
  D_{min}(X_{I},X_{J}) = \min_{i \in I} \, \min_{j \in J} \, D(X_{i},X_{j}).
\end{displaymath}
Table~2 shows the values for the parameters $\alpha_{ij}$, $\beta_{ii'}$, $\beta_{jj'}$, $\gamma_{ij}$ and $\delta$ which determine the clustering method computed by formula~(\ref{formula:new}). They are all gathered from the respective formulae~(\ref{single3}), (\ref{complete2}), (\ref{average:unweighted2}), (\ref{average:weighted}), (\ref{centroid:unweighted3}), (\ref{centroid:weighted}) and (\ref{joint2}), derived in the previous section.

\section{Glamorganshire Soils Example}
We show here a real example which has been studied by Morgan and Ray (1995) using the complete linkage method. It is the \textit{Glamorganshire soils} example, formed by similarity data between 23 different soils. A table with the similarities can be found also in Morgan and Ray (1995), where the values are given with an accuracy of three decimal places. In order to work with dissimilarities, first of all we have transformed the similarities $s(x_{i},x_{j})$ into the corresponding dissimilarities $d(x_{i},x_{j})=1-s(x_{i},x_{j})$.

The original data present a tied value for pairs of soils (3,15) and (3,20), which is responsible for two different dendrograms using the complete linkage strategy. We show them in Figures~5 and~6. Morgan and Ray (1995) explain that the 23 soils have been categorized into eight ``great soil groups'' by a surveyor. Focusing on soils 1, 2, 6, 12 and 13, which are the only members of the brown earths soil group, we see that the dendrogram in Figure~5 does not place them in the same cluster until they join soils from five other soil groups, forming the cluster (1, 2, 3, 20, 12, 13, 15, 5, 6, 8, 14, 18). From this point of view, the dendrogram in Figure~6 is better, since the corresponding cluster loses soils 8, 14 and 18, each representing a different soil group. So, in this case, we have two possible solution dendrograms and the probability of obtaining the ``good'' one is, hence, 50\%.

On the other hand, in Figure~7 we can see the multidendrogram corresponding to the \textit{Glamorganshire soils} data. The existence of a tie comprising soils 3, 15 and 20 is clear from this tree representation. Besides, the multidendrogram gives us the good classification, that is, the one with soils 8, 14 and 18 out of the brown earths soil group. Except for the internal structure of the cluster (1, 2, 3, 15, 20), the rest of the multidendrogram hierarchy coincides with that of the dendrogram shown in Figure~6.

Finally, notice that the incidence of ties depends on the accuracy with which proximity values are available. In this example, if dissimilarities had been measured to four decimal places, then the tie causing the non-unique complete linkage dendrogram might have disappeared. On the contrary, the probability of ties is higher if lower accuracy data are used. For instance, when we consider the same soils data but with an accuracy of only two decimal places, we obtain the multidendrogram shown in Figure~8, where three different ties can be observed.

\section{Conclusions}
The non-uniqueness problem in agglomerative hierarchical clustering generates several hierarchical classifications from a unique set of tied proximity data. In such cases, selecting a unique classification can be misleading. This problem has traditionally been dealt with distinct criteria, which mostly consist of the selection of one out of various resulting hierarchies. In this article we have proposed a variable-group algorithm for agglomerative hierarchical clustering that solves the ties in proximity problem. The output of this algorithm is a uniquely determined type of valued tree, which we call a multivalued tree, while graphically we represent it with a multidendrogram.

In addition we have generalized the definition of distance between clusters for the most commonly used agglomerative hierarchical methods, in order to be able to compute them using the variable-group algorithm. We have also given the corresponding generalization of Lance and Williams' formula, which enables us to get agglomerative hierarchical classifications in a recursive way. Finally, we have showed the possible usefulness of our proposal with some results obtained using data from a real example.

Gathering up the main advantages of our new proposal, we can state the following points:
\begin{itemize}
  \item When there are no ties, the variable-group algorithm gives the same result as the pair-group one.
  \item The new algorithm always gives a uniquely determined solution.
  \item In the multidendrogram representation for the results one can explicitly observe the occurrence of ties during the agglomerative process. Furthermore, the height of any fusion interval indicates the degree of heterogeneity inside the corresponding cluster.
  \item When ties exist, the variable-group algorithm is computationally more efficient than obtaining all the possible solutions following out the various ties with the pair-group alternative.
  \item The new proposal can be also computed in a recursive way using a generalization of Lance and Williams' formula.
\end{itemize}

Although ties need not be present in the initial proximity data, they may arise during the agglomeration process. For this reason and given that the results of the variable-group algorithm coincide with those of the pair-group algorithm when there are not any ties, we recommend to use directly the variable-group option. With a single action one knows whether ties exist or not, and additionally the subsequent solution is obtained.

\section*{Acknowledgments}
The authors thank A.\ Arenas for discussion and helpful comments. This work was partially supported by DGES of the Spanish Government Project No.~FIS2006--13321--C02--02 and by a grant of Universitat Rovira i Virgili.

\appendix
\section{Appendix: Proofs}
\subsection{Proof for the Unweighted Centroid Method}
Given a cluster $X_{i}$, its centroid is
\begin{displaymath}
  \overline{x}_{i} = \frac{1}{|X_{i}|} \sum_{x \in X_{i}} x,
\end{displaymath}
and the centroid of a supercluster $X_{I}$ can be expressed in terms of its constituent centroids by the equation:
\begin{equation}
  \label{proof:centroid1}
  \overline{x}_{I} = \frac{1}{|X_{I}|} \sum_{i \in I} |X_{i}| \overline{x}_{i}.
\end{equation}
Now, given two superclusters $X_{I}$ and $X_{J}$, the distance between them defined in equation~(\ref{centroid:unweighted2}) is
\begin{displaymath}
  D(X_{I},X_{J}) = \|\overline{x}_{I} - \overline{x}_{J}\|^{2} = \|\overline{x}_{I}\|^{2} + \|\overline{x}_{J}\|^{2} - 2 \ip{\overline{x}_{I}}{\overline{x}_{J}},
\end{displaymath}
where $\ip{\cdot}{\cdot}$ stands for the inner product. If we substitute each centroid by its definition~(\ref{proof:centroid1}), we obtain:
\begin{eqnarray*}
  \lefteqn{ D(X_{I},X_{J}) = \frac{1}{|X_{I}|^{2}} \sum_{i \in I} \sum_{i' \in I} |X_{i}||X_{i'}| \ip{\overline{x}_{i}}{\overline{x}_{i'}} } \\
    & & + \frac{1}{|X_{J}|^{2}} \sum_{j \in J} \sum_{j' \in J} |X_{j}||X_{j'}| \ip{\overline{x}_{j}}{\overline{x}_{j'}} \\
    & & - \frac{1}{|X_{I}||X_{J}|} \sum_{i \in I} \sum_{j \in J} |X_{i}||X_{j}| 2 \ip{\overline{x}_{i}}{\overline{x}_{j}}.
\end{eqnarray*}
Now, since
\begin{displaymath}
  2 \ip{\overline{x}_{i}}{\overline{x}_{j}} = \|\overline{x}_{i}\|^{2} + \|\overline{x}_{j}\|^{2} - \|\overline{x}_{i} - \overline{x}_{j}\|^{2},
\end{displaymath}
we have that
\begin{eqnarray*}
  \lefteqn{ D(X_{I},X_{J}) = \frac{1}{|X_{I}|^{2}} \sum_{i \in I} \sum_{i' \in I} |X_{i}||X_{i'}| \ip{\overline{x}_{i}}{\overline{x}_{i'}} } \\
    & & + \frac{1}{|X_{J}|^{2}} \sum_{j \in J} \sum_{j' \in J} |X_{j}||X_{j'}| \ip{\overline{x}_{j}}{\overline{x}_{j'}} \\
    & & - \frac{1}{|X_{I}||X_{J}|} \sum_{i \in I} \sum_{j \in J} |X_{i}||X_{j}| \|\overline{x}_{i}\|^{2} - \frac{1}{|X_{I}||X_{J}|} \sum_{i \in I} \sum_{j \in J} |X_{i}||X_{j}| \|\overline{x}_{j}\|^{2} \\
    & & + \frac{1}{|X_{I}||X_{J}|} \sum_{i \in I} \sum_{j \in J} |X_{i}||X_{j}| \|\overline{x}_{i} - \overline{x}_{j}\|^{2}.
\end{eqnarray*}
But this can be rewritten as
\begin{eqnarray*}
  \lefteqn{ D(X_{I},X_{J}) = \frac{1}{|X_{I}||X_{J}|} \sum_{i \in I} \sum_{j \in J} |X_{i}||X_{j}| \|\overline{x}_{i} - \overline{x}_{j}\|^{2} } \\
    & & - \frac{1}{|X_{I}|} \sum_{i \in I} |X_{i}| \|\overline{x}_{i}\|^{2} + \frac{1}{|X_{I}|^{2}} \sum_{i \in I} \sum_{i' \in I} |X_{i}||X_{i'}| \ip{\overline{x}_{i}}{\overline{x}_{i'}} \\
    & & - \frac{1}{|X_{J}|} \sum_{j \in J} |X_{j}| \|\overline{x}_{j}\|^{2} + \frac{1}{|X_{J}|^{2}} \sum_{j \in J} \sum_{j' \in J} |X_{j}||X_{j'}| \ip{\overline{x}_{j}}{\overline{x}_{j'}},
\end{eqnarray*}
and, grouping terms,
\begin{eqnarray*}
  \lefteqn{ D(X_{I},X_{J}) = \frac{1}{|X_{I}||X_{J}|} \sum_{i \in I} \sum_{j \in J} |X_{i}||X_{j}| \|\overline{x}_{i} - \overline{x}_{j}\|^{2} } \\
    & & - \frac{1}{|X_{I}|^{2}} \sum_{i \in I} \sum_{i' \in I} |X_{i}||X_{i'}| \left(\|\overline{x}_{i}\|^{2} - \ip{\overline{x}_{i}}{\overline{x}_{i'}}\right) \\
    & & - \frac{1}{|X_{J}|^{2}} \sum_{j \in J} \sum_{j' \in J} |X_{j}||X_{j'}| \left(\|\overline{x}_{j}\|^{2} - \ip{\overline{x}_{j}}{\overline{x}_{j'}}\right).
\end{eqnarray*}
The second and third terms can be simplified a little more, thanks to the equality
\begin{eqnarray*}
  \lefteqn{ \sum_{i \in I} \sum_{i' \in I} |X_{i}||X_{i'}| \left(\|\overline{x}_{i}\|^{2} - \ip{\overline{x}_{i}}{\overline{x}_{i'}}\right) = } \\
    & = & \sum_{i \in I} \sum_{\substack{i' \in I \\ i'>i}} |X_{i}||X_{i'}| \left(\|\overline{x}_{i}\|^{2} + \|\overline{x}_{i'}\|^{2} - 2 \ip{\overline{x}_{i}}{\overline{x}_{i'}}\right).
\end{eqnarray*}

With this simplification, we have that
\begin{eqnarray*}
  \lefteqn{ D(X_{I},X_{J}) = \frac{1}{|X_{I}||X_{J}|} \sum_{i \in I} \sum_{j \in J} |X_{i}||X_{j}| \|\overline{x}_{i} - \overline{x}_{j}\|^{2} } \\
    & & - \frac{1}{|X_{I}|^{2}} \sum_{i \in I} \sum_{\substack{i' \in I \\ i'>i}} |X_{i}||X_{i'}| \|\overline{x}_{i} - \overline{x}_{i'}\|^{2} \\
    & & - \frac{1}{|X_{J}|^{2}} \sum_{j \in J} \sum_{\substack{j' \in J \\ j'>j}} |X_{j}||X_{j'}| \|\overline{x}_{j} - \overline{x}_{j'}\|^{2},
\end{eqnarray*}
and, recalling the definition of distance between two clusters given in equation~(\ref{centroid:unweighted1}), we finally obtain the desired form of equation~(\ref{centroid:unweighted3}).

\subsection{Proof for the Joint Between-Within Method}
We give here a proof based on that of Sz\'ekely and Rizzo (2005) for their agglomerative hierarchical formulation. Using the following constants:
\begin{eqnarray}
  \label{proof:joint1}
  \theta_{ij} & = & \frac{1}{|X_{i}||X_{j}|} \sum_{x \in X_{i}} \sum_{y \in X_{j}} \|x-y\|^{\alpha}, \nonumber \\
  \theta_{ii} & = & \frac{1}{|X_{i}|^{2}} \sum_{x \in X_{i}} \sum_{x' \in X_{i}} \|x-x'\|^{\alpha},
\end{eqnarray}
the definition~(\ref{joint1}) of distance between two clusters $X_{i}$ and $X_{j}$ is
\begin{displaymath}
  D(X_{i},X_{j}) = \frac{|X_{i}||X_{j}|}{|X_{i}|+|X_{j}|} (2 \theta_{ij} - \theta_{ii} - \theta_{jj}).
\end{displaymath}
Consider now the superclusters $X_{I}$ and $X_{J}$ formed by merging clusters $X_{i}$, for all $i \in I$, and $X_{j}$, for all $j \in J$. Define the corresponding constants:
\begin{eqnarray*}
  \theta_{IJ} & = & \frac{1}{|X_{I}||X_{J}|} \sum_{x \in X_{I}} \sum_{y \in X_{J}} \|x-y\|^{\alpha} \\
    & = & \frac{1}{|X_{I}||X_{J}|} \sum_{i \in I} \sum_{j \in J} \sum_{x \in X_{i}} \sum_{y \in X_{j}} \|x-y\|^{\alpha}, \\
  \theta_{II} & = & \frac{1}{|X_{I}|^{2}} \sum_{x \in X_{I}} \sum_{x' \in X_{I}} \|x-x'\|^{\alpha} \\
    & = & \frac{1}{|X_{I}|^{2}} \sum_{i \in I} \sum_{i' \in I} \sum_{x \in X_{i}} \sum_{x' \in X_{i'}} \|x-x'\|^{\alpha} \\
    & = & \frac{1}{|X_{I}|^{2}} \sum_{i \in I} \biggl( \sum_{x \in X_{i}} \sum_{x' \in X_{i}} \|x-x'\|^{\alpha} + 2 \sum_{\substack{i' \in I \\ i'>i}} \sum_{x \in X_{i}} \sum_{x' \in X_{i'}} \|x-x'\|^{\alpha} \biggr),
\end{eqnarray*}
so that in terms of the original constants~(\ref{proof:joint1}) we have:
\begin{eqnarray*}
  \theta_{IJ} & = & \frac{1}{|X_{I}||X_{J}|} \sum_{i \in I} \sum_{j \in J} |X_{i}||X_{j}| \theta_{ij}, \\
  \theta_{II} & = & \frac{1}{|X_{I}|^{2}} \sum_{i \in I} \biggl( |X_{i}|^{2} \theta_{ii} + 2 \sum_{\substack{i' \in I \\ i'>i}} |X_{i}||X_{i'}| \theta_{ii'} \biggr).
\end{eqnarray*}
Therefore, the distance between superclusters $X_{I}$ and $X_{J}$ is given by
\begin{eqnarray*}
  \lefteqn{ D(X_{I},X_{J}) = \frac{|X_{I}||X_{J}|}{|X_{I}|+|X_{J}|} (2 \theta_{IJ} - \theta_{II} - \theta_{JJ}) } \\
                 & = & \frac{|X_{I}||X_{J}|}{|X_{I}|+|X_{J}|} \biggl[ \frac{2}{|X_{I}||X_{J}|} \sum_{i \in I} \sum_{j \in J} |X_{i}||X_{j}| \theta_{ij} \\
    & & - \frac{1}{|X_{I}|^{2}} \sum_{i \in I} \biggl( |X_{i}|^{2} \theta_{ii} + 2 \sum_{\substack{i' \in I \\ i'>i}} |X_{i}||X_{i'}| \theta_{ii'} \biggr) \\
    & & - \frac{1}{|X_{J}|^{2}} \sum_{j \in J} \biggl( |X_{j}|^{2} \theta_{jj} + 2 \sum_{\substack{j' \in J \\ j'>j}} |X_{j}||X_{j'}| \theta_{jj'} \biggr) \biggr].
\end{eqnarray*}
Simplify
\begin{eqnarray*}
  \lefteqn{ \sum_{i \in I} \biggl( |X_{i}|^{2} \theta_{ii} + 2 \sum_{\substack{i' \in I \\ i'>i}} |X_{i}||X_{i'}| \theta_{ii'} \biggr) = } \\
    & = & \sum_{i \in I} \biggl[ |X_{i}|^{2} \theta_{ii} + \sum_{\substack{i' \in I \\ i'>i}} |X_{i}||X_{i'}| (2 \theta_{ii'} - \theta_{ii} - \theta_{i'i'} + \theta_{ii} + \theta_{i'i'}) \biggr] \\
    & = & \sum_{i \in I} \biggl[ |X_{i}|^{2} \theta_{ii} + \sum_{\substack{i' \in I \\ i'>i}} |X_{i}||X_{i'}| (\theta_{ii} + \theta_{i'i'}) + \sum_{\substack{i' \in I \\ i'>i}} (|X_{i}|+|X_{i'}|) D(X_{i},X_{i'}) \biggr] \\
    & = & |X_{I}| \sum_{i \in I} |X_{i}| \theta_{ii} + \sum_{i \in I} \sum_{\substack{i' \in I \\ i'>i}} (|X_{i}|+|X_{i'}|) D(X_{i},X_{i'}),
\end{eqnarray*}
where in last equality we have used the equivalence
\begin{displaymath}
  \sum_{i \in I} \biggl[ |X_{i}|^{2} \theta_{ii} + \sum_{\substack{i' \in I \\ i'>i}} |X_{i}||X_{i'}| (\theta_{ii} + \theta_{i'i'}) \biggr] = |X_{I}| \sum_{i \in I} |X_{i}| \theta_{ii}.
\end{displaymath}
Hence,
\begin{eqnarray*}
  \lefteqn{ (|X_{I}|+|X_{J}|)D(X_{I},X_{J}) = 2 \sum_{i \in I} \sum_{j \in J} |X_{i}||X_{j}| \theta_{ij} } \\
    & & - |X_{J}| \sum_{i \in I} |X_{i}| \theta_{ii} - \frac{|X_{J}|}{|X_{I}|} \sum_{i \in I} \sum_{\substack{i' \in I \\ i'>i}} (|X_{i}|+|X_{i'}|) D(X_{i},X_{i'}) \\
    & & - |X_{I}| \sum_{j \in J} |X_{j}| \theta_{jj} - \frac{|X_{I}|}{|X_{J}|} \sum_{j \in J} \sum_{\substack{j' \in J \\ j'>j}} (|X_{j}|+|X_{j'}|) D(X_{j},X_{j'}),
\end{eqnarray*}
or, equivalently,
\begin{eqnarray*}
  \lefteqn{ (|X_{I}|+|X_{J}|)D(X_{I},X_{J}) = \sum_{i \in I} \sum_{j \in J} |X_{i}||X_{j}| 2 \theta_{ij} } \\
    & & - \sum_{i \in I} |X_{i}| \theta_{ii} \sum_{j \in J} |X_{j}| - \sum_{i \in I} |X_{i}| \sum_{j \in J} |X_{j}| \theta_{jj} \\
    & & - \frac{|X_{J}|}{|X_{I}|} \sum_{i \in I} \sum_{\substack{i' \in I \\ i'>i}} (|X_{i}|+|X_{i'}|) D(X_{i},X_{i'}) \\
    & & - \frac{|X_{I}|}{|X_{J}|} \sum_{j \in J} \sum_{\substack{j' \in J \\ j'>j}} (|X_{j}|+|X_{j'}|) D(X_{j},X_{j'}),
\end{eqnarray*}
which is also the same as
\begin{eqnarray*}
  \lefteqn{ (|X_{I}|+|X_{J}|)D(X_{I},X_{J}) = \sum_{i \in I} \sum_{j \in J} (|X_{i}|+|X_{j}|) D(X_{i},X_{j}) } \\
    & & - \frac{|X_{J}|}{|X_{I}|} \sum_{i \in I} \sum_{\substack{i' \in I \\ i'>i}} (|X_{i}|+|X_{i'}|) D(X_{i},X_{i'}) \\
    & & - \frac{|X_{I}|}{|X_{J}|} \sum_{j \in J} \sum_{\substack{j' \in J \\ j'>j}} (|X_{j}|+|X_{j'}|) D(X_{j},X_{j'}).
\end{eqnarray*}
And this is exactly the desired formulation given in equation~(\ref{joint2}).

\clearpage
\begin{center}
  \small
  \begin{tabular}{|lcccc|}
    \multicolumn{5}{c}{\footnotesize Table 1. Parameter Values for the Lance and Williams' Formula} \\
    \hline
    Method               & $\alpha_{i}$ & $\alpha_{i'}$ & $\beta$ & $\gamma$ \\
    \hline
    \hline
    Single linkage       & $\frac{1}{2}$ & $\frac{1}{2}$ & $0$ & $-\frac{1}{2}$ \\[1mm]
    Complete linkage     & $\frac{1}{2}$ & $\frac{1}{2}$ & $0$ & $+\frac{1}{2}$ \\[1mm]
    Unweighted average   & $\frac{|X_{i}|}{|X_{i}|+|X_{i'}|}$ & $\frac{|X_{i'}|}{|X_{i}|+|X_{i'}|}$ & $0$ & $0$ \\[1mm]
    Weighted average     & $\frac{1}{2}$ & $\frac{1}{2}$ & $0$ & $0$ \\[1mm]
    Unweighted centroid  & $\frac{|X_{i}|}{|X_{i}|+|X_{i'}|}$ & $\frac{|X_{i'}|}{|X_{i}|+|X_{i'}|}$ & $-\frac{|X_{i}||X_{i'}|}{(|X_{i}|+|X_{i'}|)^{2}}$ & $0$ \\[1mm]
    Weighted centroid    & $\frac{1}{2}$ & $\frac{1}{2}$ & $-\frac{1}{4}$ & $0$ \\[1mm]
    Joint between-within & $\frac{|X_{i}|+|X_{j}|}{|X_{i}|+|X_{i'}|+|X_{j}|}$ & $\frac{|X_{i'}|+|X_{j}|}{|X_{i}|+|X_{i'}|+|X_{j}|}$ & $-\frac{|X_{j}|}{|X_{i}|+|X_{i'}|+|X_{j}|}$ & $0$ \\[1mm]
    \hline
  \end{tabular}
\end{center}

\clearpage
\begin{center}
  \small
  \begin{tabular}{|lccccc|}
    \multicolumn{6}{c}{\footnotesize Table 2. Parameter Values for the Variable-Group Formula} \\
    \hline
    Method               & $\alpha_{ij}$ & $\beta_{ii'}$ & $\beta_{jj'}$ & $\gamma_{ij}$ & $\delta$ \\
    \hline
    \hline
    Single linkage       & $\frac{1}{|I||J|}$ & $0$ & $0$ & $\frac{1}{|I||J|}$ & $0$ \\[1mm]
    Complete linkage     & $\frac{1}{|I||J|}$ & $0$ & $0$ & $\frac{1}{|I||J|}$ & $1$ \\[1mm]
    Unweighted average   & $\frac{|X_{i}||X_{j}|}{|X_{I}||X_{J}|}$ & $0$ & $0$ & $0$ & $-$ \\[1mm]
    Weighted average     & $\frac{1}{|I||J|}$ & $0$ & $0$ & $0$ & $-$ \\[1mm]
    Unweighted centroid  & $\frac{|X_{i}||X_{j}|}{|X_{I}||X_{J}|}$ & $-\frac{|X_{i}||X_{i'}|}{|X_{I}|^{2}}$ & $-\frac{|X_{j}||X_{j'}|}{|X_{J}|^{2}}$ & $0$ & $-$ \\[1mm]
    Weighted centroid    & $\frac{1}{|I||J|}$ & $-\frac{1}{|I|^{2}}$ & $-\frac{1}{|J|^{2}}$ & $0$    & $-$ \\[1mm]
    Joint between-within & $\frac{|X_{i}|+|X_{j}|}{|X_{I}|+|X_{J}|}$ & $-\frac{|X_{J}|}{|X_{I}|} \frac{|X_{i}|+|X_{i'}|}{|X_{I}|+|X_{J}|}$ & $-\frac{|X_{I}|}{|X_{J}|} \frac{|X_{j}|+|X_{j'}|}{|X_{I}|+|X_{J}|}$ & $0$ & $-$ \\[1mm]
    \hline
  \end{tabular}
\end{center}

%

\clearpage
\begin{center}
  \begin{picture}(120,90)
    \put(  0, 0){\framebox(120,90){}}
    \put( 20,70){\circle{20}}
    \put( 60,70){\circle{20}}
    \put(100,70){\circle{20}}
    \put(100,20){\circle{20}}
    \put( 30,70){\line( 1, 0){20}}
    \put( 70,70){\line( 1, 0){20}}
    \put(100,30){\line( 0, 1){30}}
    \put( 20,70){\makebox(0,0){$x_{1}$}}
    \put( 60,70){\makebox(0,0){$x_{2}$}}
    \put(100,70){\makebox(0,0){$x_{3}$}}
    \put(100,20){\makebox(0,0){$x_{4}$}}
    \put( 40,68){\makebox(0,0)[t]{2}}
    \put( 80,68){\makebox(0,0)[t]{2}}
    \put( 99,45){\makebox(0,0)[r]{3}}
  \end{picture} \\
  {\footnotesize Figure 1. Toy Graph with Four Individuals and Shortest Path Distances}
\end{center}

\clearpage
\begin{center}
  \makebox{\includegraphics{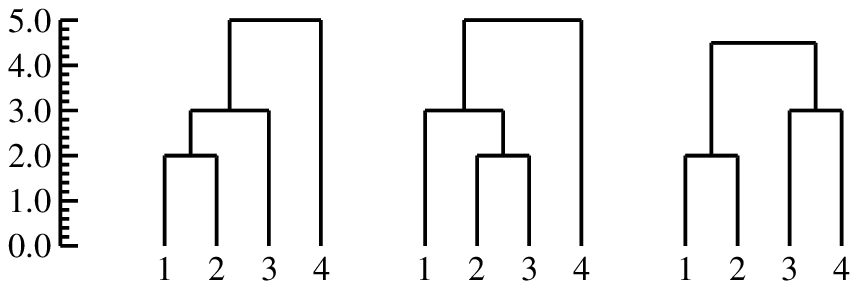}} \\
  {\footnotesize Figure 2. Unweighted Average Dendrograms for the Toy Example}
\end{center}

\clearpage
\begin{center}
  \makebox{\includegraphics{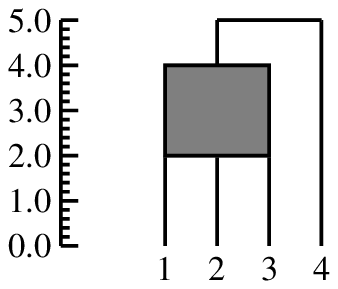}} \\
  {\footnotesize Figure 3. Unweighted Average Multidendrogram for the Toy Example}
\end{center}

\clearpage
\begin{center}
  \begin{picture}(170,130)
    \put(  0,  0){\framebox(170,130){}}
    \put( 20,110){\circle{20}}
    \put( 20, 60){\circle{20}}
    \put( 20, 20){\circle{20}}
    \put( 60, 20){\circle{20}}
    \put(110,110){\circle{20}}
    \put(110, 60){\circle{20}}
    \put(130, 30){\circle{20}}
    \put(150,110){\circle{20}}
    \put(150, 60){\circle{20}}
    \put( 20, 30){\line( 0, 1){20}}
    \put( 30, 20){\line( 1, 0){20}}
    \put(120,110){\line( 1, 0){20}}
    \put(120, 60){\line( 1, 0){20}}
    \put(116, 52){\line( 2,-3){9}}
    \put(144, 52){\line(-2,-3){9}}
    \put( 20,110){\makebox(0,0){$X_{1}$}}
    \put(110,110){\makebox(0,0){$X_{2}$}}
    \put(150,110){\makebox(0,0){$X_{3}$}}
    \put( 20, 60){\makebox(0,0){$X_{4}$}}
    \put( 20, 20){\makebox(0,0){$X_{5}$}}
    \put( 60, 20){\makebox(0,0){$X_{6}$}}
    \put(110, 60){\makebox(0,0){$X_{7}$}}
    \put(150, 60){\makebox(0,0){$X_{8}$}}
    \put(130, 30){\makebox(0,0){$X_{9}$}}
  \end{picture} \\
  {\footnotesize Figure 4. Simultaneous Occurrence of Different Superclusters}
\end{center}

\clearpage
\begin{center}
  \makebox{\includegraphics{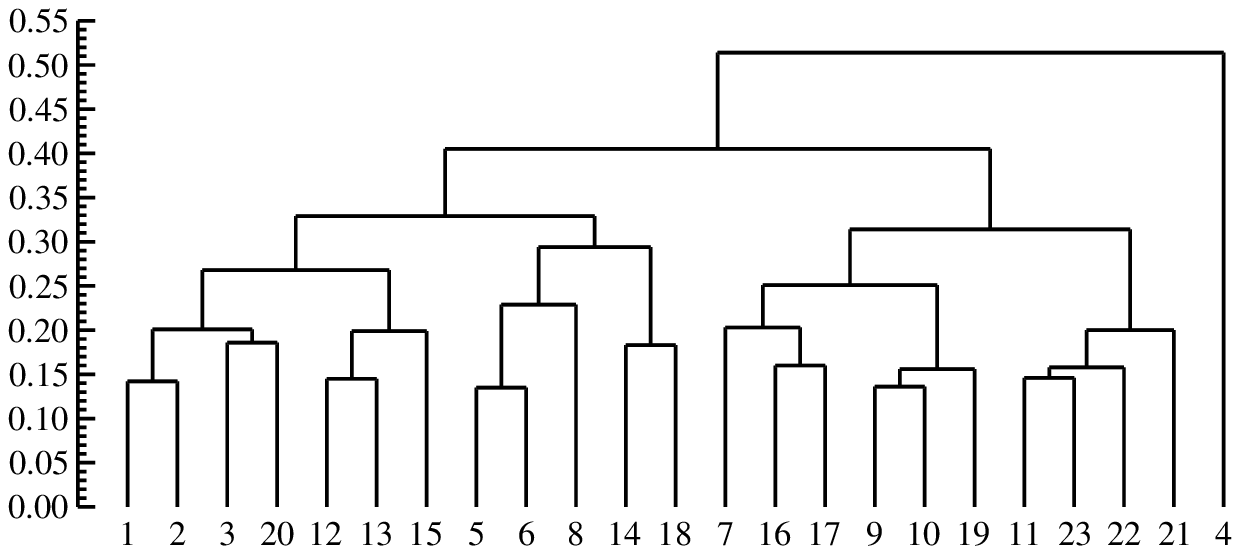}} \\
  {\footnotesize Figure 5. First Complete Linkage Dendrogram for the Soils Data}
\end{center}

\clearpage
\begin{center}
  \makebox{\includegraphics{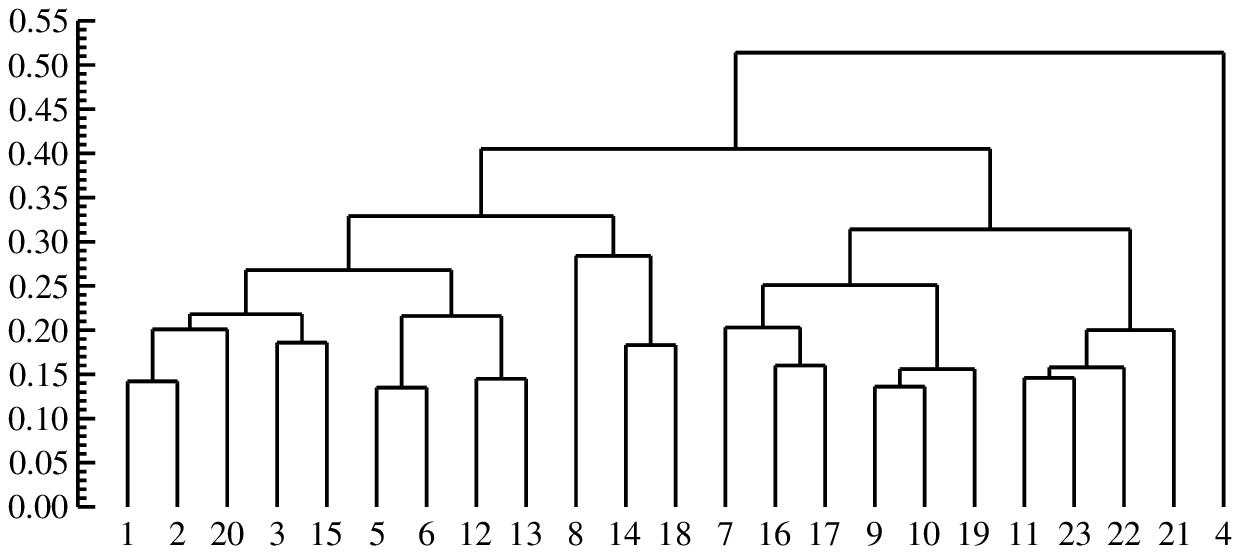}} \\
  {\footnotesize Figure 6. Second Complete Linkage Dendrogram for the Soils Data}
\end{center}

\clearpage
\begin{center}
  \makebox{\includegraphics{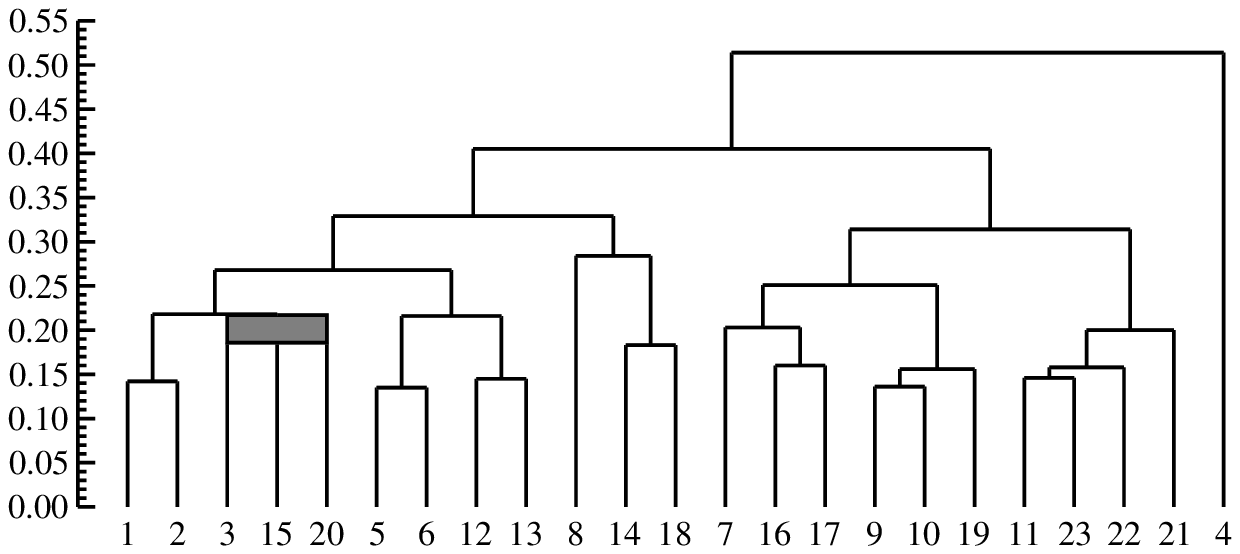}} \\
  {\footnotesize Figure 7. Complete Linkage Multidendrogram for the Soils Data}
\end{center}

\clearpage
\begin{center}
  \makebox{\includegraphics{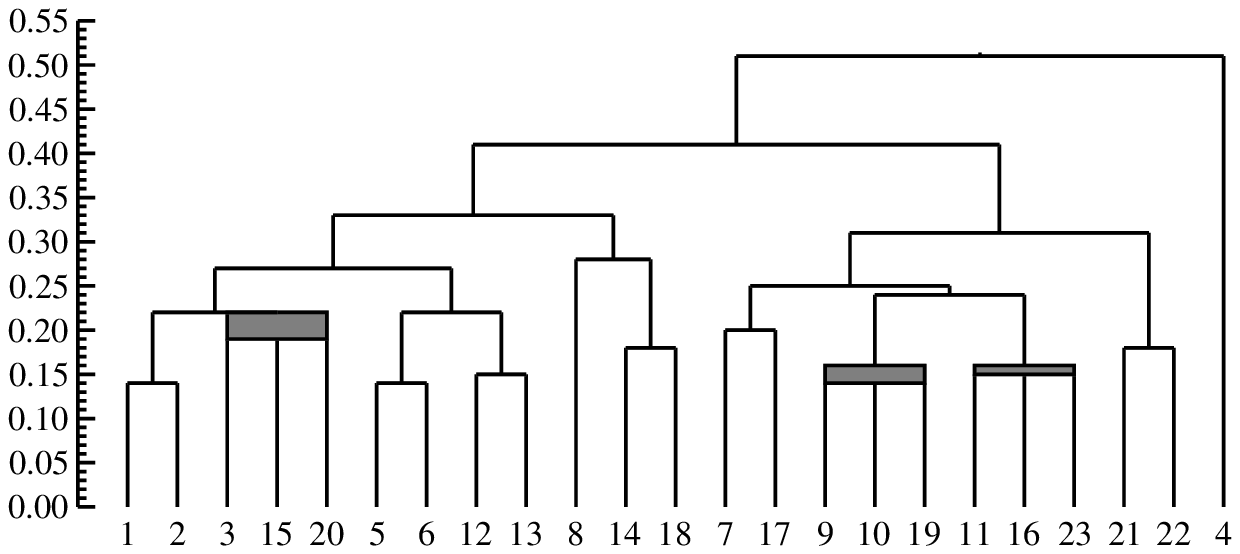}} \\
  \raggedright {\footnotesize Figure 8. Complete Linkage Multidendrogram for the Soils Data with an Accuracy of Two Decimal Places}
\end{center}

\end{document}